\documentclass[preprints,article,accept,moreauthors,pdftex]{Definitions/mdpi}

\newcommand{\beq}{\begin{equation}}
\newcommand{\eeq}{\end{equation}}
\newcommand{\bea}{\begin{eqnarray}}
\newcommand{\eea}{\end{eqnarray}}

\firstpage{1} 
\pubvolume{1}
\issuenum{1}
\articlenumber{0}
\pubyear{2021}
\copyrightyear{2020}
\datereceived{} 
\dateaccepted{} 
\datepublished{} 
\hreflink{https://doi.org/} 
\pdfoutput=1

\usepackage{amsmath}
\usepackage{amssymb}
\usepackage{amsbsy}
\usepackage{amsfonts}
\usepackage{amsopn}
\usepackage{amstext}
\usepackage{graphicx}
\usepackage{amssymb}
\usepackage{amsfonts}
\usepackage{amsmath}
\usepackage{graphicx}
\usepackage[english]{babel}
\usepackage{color}
\usepackage{slashed}
\usepackage{esint}
\usepackage[dvips]{epsfig}
\usepackage[dvips]{graphicx}
\usepackage{float}
\usepackage{natbib}
\usepackage{units}
\usepackage{textcomp}
\usepackage{wasysym}

\Title{Schwarzschild-like Wormholes in Asymptotically Safe Gravity}

\TitleCitation{Title}

\Author{G. Alencar $^{1,2,}*$  and M.Nilton $^{1}$}

\AuthorNames{Firstname Lastname, Firstname Lastname and Firstname Lastname}

\AuthorCitation{Lastname, F.; Lastname, F.; Lastname, F.}

\address{%
$^{1}$ \quad Universidade Federal do Cear\'a, Fortaleza-CE, Brazil\\
$^{2}$ \quad International Institute of Physics - Federal University of Rio Grande do Norte, Campus Universit\'ario, Lagoa Nova, Natal, RN 59078-970, Brazil; matheus.nilton@fisica.ufc.br}

\corres{Correspondence: geova@fisica.ufc.br}

\abstract{In this paper, we analyze the Schwarzschild-like wormhole in the Asymptotically Safe Gravity(ASG) scenario. The ASG corrections are implemented via renormalization group methods, which, as consequence, provides a new tensor $X_{\mu\nu}$ as a source to improved field equations, and promotes the Newton's constant into a running coupling constant. In particular, we check whether the radial energy conditions are satisfied and compare with the results obtained from the usual theory. We show that only in the particular case of the wormhole being asymptotically flat(Schwarzschild Wormholes) that the radial energy conditions are satisfied at the throat, depending on the chosen values for its radius $r_0$. In contrast, in the general Schwarzschild-like case, there is no possibility of the energy conditions being satisfied nearby the throat, as in the usual case. After that, we calculate the radial state parameter, $\omega(r)$, in $r_0$, in order to verify what type of cosmologic matter is allowed at the wormhole throat, and we show that in both cases there is the possibility of the presence of exotic matter, phantom or quintessence-like matter. Finally, we give the $\omega(r)$ solutions for all regions of space. Interestingly, we find that Schwarzschild-like Wormholes with excess of solid angle of the sphere in the asymptotic limit have the possibility of having non-exotic matter as source for certain values of the radial coordinate $r$. Furthermore, it was observed that quantum gravity corrections due the ASG necessarily imply regions with phantom-like matter, both for Schwarzschild and for Schwarzschild-like wormholes. This reinforces the supposition that a phantom fluid is always present for wormholes in this context.}


\begin{document}
\section{Introduction}
The Einstein's theory of general relativity predicts the existence of interesting objects that serve as tunnels connecting regions of spacetime that are asymptotically flat. This type of solution was first found by Einstein and Rosen\cite{Einstein:1935tc}, which were later called Wormholes by Misner and Wheeler\cite{Misner:1957mt}. However, the first traversable Wormhole solution was given by Morris-Thorne solution\cite{Morris:1988cz}, which is represented by a static spherical symmetric solution of the form:

\begin{equation}
ds^{2} = e^{2\Phi(r)}dt^2-\frac{dr^2}{1-b(r)/r}-r^2d\Omega_2,
\end{equation}
where $d\Omega_2$ is the 2-sphere line element, $e^{2\Phi(r)}$ is the redshift function and $b(r)$ the shape function. In order to ensure traversability, some restrictions are imposed on these functions\cite{Morris:1988cz,Cataldo:2015vra}, such as the non-existence of an event horizon, the existence of a minimum value for $b(r)$, which is for some $r=r_0$, with $b(r_0)=r_0$, characterizing the wormhole throat region, and the flare-out condition, represented by the relation $b'(r_0)<1$, where ' is the differentiation with respect to radial coordinate $r$. However, it can be shown that for a wormhole to be traversable, at least in the context of General Relativity, it must necessarily contain matter that violates at least one of the the energy conditions\cite{Bronnikov:2014bda,Visser:1995cc,Blazquez-Salcedo:2020czn,Konoplya:2021hsm}, being possible to have wormholes with phantom energy as a source \cite{Sushkov:2005kj,Lobo:2005us} or even Casimir energy\cite{Garattini:2019ivd,Jusufi:2020rpw,Alencar:2021ejd,Oliveira:2021ypz,Carvalho:2021ajy}.

Wormholes that have geometries that are asymptotically flat are of particular interest \cite{Morris:1988cz}, although more general cases have already been studied in the literature \cite{Lemos:2004vs,Barcelo:1995gz,Lemos:2003jb}. The Schwarzschild's solution can be thought of as an example of an asymptotically flat wormhole, with the shape-function $b(r)=r_0=\textrm{constant}$ and the redshift function with the form $e^{2\Phi(r)}=1-r_0/r$. However, this would not be a traversable wormhole, as it would have an event horizon in $r=r_0$. In order to construct a traversable Schwarzschild Wormhole we must maintain the radial component of the metric and require that the redshift function has no horizons. We can also construct a generalized version of the Schwarzschild Wormhole, considering a linear shape function with the form $b(r)=(1-\beta)r_0+\beta r$, where $\beta$ is a constant parameter and the $\beta=0$ being the particular case of the Schwarzschild wormhole, also being considered a particular case of a self-dual wormhole with a null energy density \cite{Dadhich:2001fu,Cataldo:2002jw}. A consequence of the $\beta$ parameter is the presence of a increase or deficit of the solid angle  at the asymptotic limit, depending on whether $\beta$ is positive or negative. This wormhole was called Schwarzschild-like and their properties was studied in \cite{Cataldo:2017ard}. Although Schwarzschild-like Wormholes are traversable, the necessity for exotic matter is again seen.

We can question whether, when considering quantum effects, there is the possibility of a wormhole being formed with non-exotic matter obeying the energy conditions. However, a quantum theory of gravity has not yet been well established. One of the methods found in the literature to consider quantum effects is to treat the gravitational field as a quantum field that is asymptotically safe. This method is particularly interesting as it is UV complete  \cite{as}. The guarantee of the existence of a fixed point for the gravity renormalization group flow is confirmed by various methods that have been applied in different models\cite{Reuter:1996cp,Lauscher:2001ya,Litim:2003vp,Machado:2007ea,Benedetti:2009rx,Manrique:2011jc,Christiansen:2012rx,Morris:2015oca,Demmel:2015oqa,Platania:2017djo,Christiansen:2017bsy,Falls:2018ylp,Narain:2009fy,Oda:2015sma,Eichhorn:2017ylw,Eichhorn:2019yzm,Reichert:2019car,Daas:2020dyo,Bonanno:2015fga,Bonanno:2016dyv,Bonanno:2017gji,Bonanno:2018gck,Platania:2019qvo,Platania:2020lqb,Bonanno:2017zen,Adeifeoba:2018ydh,Platania:2019kyx}.This method is based on solving the exact group renormalization equation in order to obtain the effective average action $\Gamma_{k}$ \cite{Wetterich:1989xg} 
\begin{equation}
\label{ERGE}
k\partial_{k}\Gamma_{k}=\frac{1}{2}\textrm{Tr}\left[(\Gamma^{(2)}_{k}+\mathcal{R}_{k})^{-1}k\partial_{k}\mathcal{R}_{k}\right],
\end{equation}
where $\Gamma^{(2)}_{k}$ is the Hessian of $\Gamma_{k}$ and the $\mathcal{R}_{k}$ is the IR-cutoff, which is required to be a quadratic function of the momentum $k$ \cite{Reuter:1996cp}.

In the case of the gravitational field, the effective average action is a functional which depends on the metric and the parameter $k$, a variable responsible for introducing the infrared (IR) cutoff. The functional $\Gamma_{k}[g_{\mu\nu}]$ describes the quantum fluctuations in the gravitational field for momentum scales of the order of $k$. The issue with this method is the difficult to solve the exact renormalization group equation \eqref{ERGE}. However, what we can do is consider the ``Einstein-Hilbert truncation''\cite{Reuter:1996cp,Dou:1997fg,Lauscher:2001ya,Reuter:2001ag,Souma:1999at,Percacci:2003jz,Reuter:2002kd}, which consists to consider the projection of the renormalization group flow on a subspace with finite dimension which that takes into account the essential physics, and so, an expansion of $\Gamma_{k}$ in the basis of this spanned space, $\sqrt{g}$, $\sqrt{g}R$, is done.  The coefficients of the expansion of $\Gamma_{k}$ contains the running coupling constant $G(k)$ which is a solution of the $\beta$-function\cite{Reuter:1996cp,Reuter:2003ca,Rodrigues:2015hba,Reuter:2004nx}.

Numerical solutions of this $\beta$-function in the infrared limit and near the fixed point provides the following form of the running coupling constant \cite{Bonanno:2000ep,Bonanno:2006eu,Bonanno:2002zb,Reuter:2004nv,Moti:2018rho}
\begin{equation}
\label{running-constant}
G(k) = \frac{G(k_{0})}{1+\omega_{q}G(k_{0})(k^{2}-k_{0}^{2})},
\end{equation}
where $k$ is a moment scale introduced in the average effect action to implement the infrared (IR) cutoff, $k_{0}$ is a reference scale and $\omega_q = \frac{4}{\pi}(1-\frac{\pi^2}{144})$. It can be shown that for a suitable choice of the reference scale we have $k_{0}=0$ with $G_{\textrm{N}} = G(k_{0}\rightarrow 0) = G_{0}$ being the Newton's constant. It can be seen that in models with flat background, the cutoff moment $k$ can be written as a function of the position with an inverse length dimension. However, this dependence on coordinates can spoil the analysis in curved spaces. To eliminate this dependence on coordinates, we can consider a improvement of the form $\omega k^{2} = \xi f(\chi)$ in curved spaces, where $\chi$ is a function of the curvature invariants and $\xi$ is a small scaling constant \cite{Moti:2018rho}. With this, the improved coupling constant \eqref{running-constant} can be written as:
\begin{equation}
\label{eq:running-coupling-constant}
G(\chi) = \frac{G_{0}}{1+f(\chi)},
\end{equation}
where $f(\chi) = \xi/\chi$ is called the anti-screening function. Therefore, the quantum corrections due the ASG theory modify the Einstein-Hilbert Action just replacing the Newton’s constant into a function of the curvature invariants given by the Eq. \eqref{eq:running-coupling-constant}, providing the following modified action\cite{Moti:2018rho}
\begin{equation}
S = \frac{1}{16\pi}\int d^{4}x\frac{\sqrt{-g}}{G(\chi)}R+\int d^{4}x\sqrt{-g}\mathcal{L}_{m}.
\end{equation}
By varying the above action with respect to the metric and using the principle of least action we obtain the quantum improved field equations\cite{Moti:2018rho}
\begin{equation}
G_{\mu\nu} = 8\pi G(\chi) T_{\mu\nu} + G(\chi) X_{\mu\nu} (\chi) \ ,\label{IEM}
\end{equation}
where $T_{\mu\nu}$ is the classical energy-moment tensor, and $X_{\mu\nu}$ is a covariant tensor that arises due the derivation of $G(\chi)$ when we vary the modified action with respect to the metric. This tensor dictates the dynamics of the scalar field $G(\chi)$, and is defined as:

\[
    X_{\mu\nu}(\chi) = \Big( \nabla_{\mu}\nabla_{\nu} - g_{\mu\nu}\square \Big) G(\chi)^{-1}   - \frac{1}{2} \bigg( R\mathcal{K}(\chi) \frac{\delta\chi}{\delta g^{\mu\nu}} +
 \]
\begin{equation}
 \partial_{\kappa}\Big (R\mathcal{K}(\chi)\frac{\partial\chi}{\partial (\partial_{\kappa}g^{\mu\nu})}\Big) + \partial_{\kappa}\partial_{\lambda}\Big (R\mathcal{K}(\chi)\frac{\partial\chi}{\partial (\partial_{\lambda}\partial_{\kappa}g^{\mu\nu})}\Big ) \bigg ) \ ,
\end{equation}
with $\mathcal{K}(\chi)\equiv  \frac{2\partial{G(\chi)}/\partial{\chi}}{G(\chi)^2}$ \cite{Moti:2019mws}.
This tensor can be thought of as the energy-moment tensor that describes the 4-momentum of the field $G(\chi)$. 

The tensor $X_{\mu\nu}$ depends on the parameter $\chi$, and expanding up to first order leads \cite{Moti:2018rho}

\begin{equation}
  X_{\mu\nu} \simeq \nabla_{\mu}\nabla_{\nu} G(\chi)^{-1} - g_{\mu\nu}\square G(\chi)^{-1} \ .
\end{equation}
As already mentioned, the parameter $\chi$ is defined in terms of curvature invariants constructed from the components of the Riemann tensor, such as $R, R_{\alpha\beta}R^{\alpha\beta}, R_{\alpha\beta\kappa\lambda}R^{\alpha\beta\kappa\lambda},\cdots$ \cite{Moti:2018rho}. Evidently, there are several possible choices for the parameter $\chi$\cite{Pawlowski:2018swz}. However, working with non-vacuum solutions suggests some simple choices such as $\chi = R^{-1}$ or $\chi = (R_{\alpha\beta}R^{\alpha\beta})^{ 1/2}$\cite{Moti:2019mws}.

In this direction, the traversability conditions for wormholes using the quantum improvement due to Asymptotically Safety in Quantum Gravity(ASQG) was investigated in \cite{Moti:2020whf}. Considering a linear equation of state they found that the Morris-Thorne solution is still traversable in this context depending on the parameter values, and, beyond that, the possibility of traversable wormholes with nonexotic matter, as long as they are pseudospherical. However, a linear equation of state rules out a large number of possible models. Furthermore, the authors restricted themselves to regions close to the throat and, therefore, did not provide information about the distribution of matter along the spacetime generated by the wormhole. In order to test this, the authors of Ref. \cite{Alencar:2021enh} studied the quantum improvement of the Ellis-Bronnikov wormhole due to ASQG. They found that, despite of the fact that nonexotic matter  is possible  at the wormhole throat, in other regions of spacetime a Phantom matter is necessary. With this, they speculate if Phantons are always necessary in such scenarios. 

Therefore, in this work we put forward the study of wormhole corrections due to ASQG. We consider the zero-tidal Schwarzschild-like wormholes in the context of the quantum improvement of gravity theory due to functional renormalization group methods to describe asymptotic safe quantum gravity. In Section II, we will present the zero-tidal Schwarzschild-like wormhole spacetime and their modified state equations due the corrections above, considering Ricci scalar, squared Ricci tensor and Kretschmann scalar as anti-screening functions, and we analyze these quantum effects for the case of Schwarzschild wormholes. We will find that in the Schwarzschild wormhole case the radial energy conditions can be satisfied at the throat, depending on its radius value $r_0$, but, we necessarily have the presence of exotic matter, as phantom or quintessence energies. In section III we have considered the same corrections but in the Schwarzschild-like wormhole case. We see that in the case of an excess of solid angle ($\beta<0$), we have the possibility of having a region with nonexotic matter depending on the values of $r_0$ and $\beta$, but, the energy conditions can not be satisfied in this case. In section IV, we conclude the paper and we give the final considerations.

\section{Schwarzschild Wormhole solution in ASG}
Every wormhole solution that is static and spherically symmetric can be described by the Morris-Thorne metric:

\begin{equation}
   d{s}^2 = e^{2\Phi(r)} d{t}^2 - \frac{d{r}^2}{1-b(r)/r} -r^2d{\Omega}_{2} \ , \label{SSM}
\end{equation}
where $d\Omega_{2}=d\theta^{2}+\sin^{2}{\theta}d\varphi^{2}$ is the line element of a 2-sphere, $e^{2\Phi(r)}$ is  the redshift function and $b(r)$ the shape function.

Considering an anisotropic fluid as source to generate the wormhole spacetime $ T^{\mu}_{\nu} = \text{Diag} [\rho(r),-p_r(r),-p_l(r),-p_l(r)] \ $, the improved field equations \eqref{IEM} lead to \cite{Moti:2020whf}
\begin{eqnarray}
    \kappa \rho  = &(1+f) \frac{b^{'}}{r^2} -  (1-\frac{b}{r}) (f''+\frac{2}{r} f')+ \frac{b^{'}r-b}{2r^2}f' \ ,\\
    \kappa p_r  = & -(1+f) \left(\frac{b}{r^3} -\frac{2\Phi^{'}}{r}(1-\frac{b}{r}) \right) + (1-\frac{b}{r}) \left( \Phi^{'}+\frac{2}{r}\right) f'  \ ,  \\
   \kappa p_l  = & -(1+f) \left( \frac{b'r-b}{2r^2} (\Phi'+\frac{1}{r} ) - (1-\frac{b}{r})(\Phi''+\Phi'^2+\frac{\Phi'}{r}) \right) \nonumber\\
    & + (1-\frac{b}{r}) \left(( \Phi^{'}+\frac{1}{r})f'+f''\right)  - \frac{b^{'}r-b}{2r^2}f' \ ,
\end{eqnarray}
where the prime denotes differentiation with respect to $r$, $\kappa=8\pi G_0$, $\rho$ is the energy density and $p_r$ and $p_l$ are the radial and lateral pressures, respectively.

In this paper we consider the case of a zero-tidal Schwarzschild-like wormhole spacetime, which is the particular case where the redshift function is equal to unity, that is, $\Phi(r)=0$, and the shape function is linear in $r$, $b(r)=\alpha + \beta r$, where $\alpha$ and $\beta$ are constants. The condition of minimum at the throat, $b(r_0)=r_0$, implies that $\alpha = (1-\beta)r_0$, leading to $b(r)=(1-\beta)r_0+\beta r$. With this, our metric becomes:
\begin{eqnarray}
\label{SLW-metric}
ds^{2} = dt^{2}-\frac{dr^{2}}{(1-\beta)\left(1-\frac{r_0}{r}\right)}-r^{2}d\Omega^{2}.
\end{eqnarray}
where $r_0$ is the throat radius and $\beta$ is a constant. The flare-out condition, $b'(r_0) \leq 1$, implies that $\beta \leq 1$. Therefore, we must have $\beta < 1$ in order to ensure that the radial component of the metric will not diverge. The particular case $\beta = 0$ provides the zero-tidal Schwarzschild wormhole solution, which is asymptotically flat. The case where $\beta \neq 0$ characterizes the zero-tidal Schwarzschild-like Wormhole, which were studied in the context of General Relativity in \cite{Cataldo:2017ard}. One of the consequences of the parameter $\beta$ is that the new asymptotic form of the metric will have a solid angle deficit, if $0<\beta<1$, or an excess of solid angle if $\beta < 0$. With the metric \eqref{SLW-metric}, our equations becomes
\begin{eqnarray}
\kappa\rho	&=&(1+f)\frac{\beta}{r^{2}}-(1-\beta)\left(1-\frac{r_{0}}{r}\right)\left(f''+\frac{2}{r}f'\right)-\frac{(1-\beta)r_{0}}{2r^{2}}f', \label{IEQ-tt-s}
\\
\kappa p_{r}&=&-(1+f)\frac{[(1-\beta)r_{0}+\beta r]}{r^{3}}+\frac{2}{r}(1-\beta)\left(1-\frac{r_{0}}{r}\right)f'\ , \label{IEQ-rr-s}
\\
\kappa p_{l}&=&(1+f)\frac{(1-\beta)r_{0}}{2r^{3}}+(1-\beta)\left(1-\frac{r_{0}}{r}\right)\left(\frac{1}{r}f'+f''\right)+\frac{(1-\beta)r_{0}}{2r^{2}}f'\ . \label{IEQ-ll-s}
\end{eqnarray}
Now we must choose the function $f=\xi/\chi$. As noted earlier, the parameter $\xi$ is a scaling constant, that in the limit $\xi \rightarrow 0$ provides the usual case of General Relativity, and $\chi$ is written in terms of the curvature invariants. We will consider models with the Ricci scalar, squared Ricci tensor and Kretschmann scalar. These models for the anti-screening function were used by the authors in \cite{Moti:2020whf} to investigate the traversability conditions of the Morris-Thorne wormhole solution in the context of Asymptotically Safe Gravity. They found that depending on the parameter values, the improved solution remains traversable but still in the presence of exotic matter. In this sense, we will use the same improvement to check if the presence of exotic matter is again seen for Schwarzschild-like wormholes in this quantum context. For the metric \eqref{SLW-metric} these quantities are given respectively by

\begin{equation}
\begin{split}
R=&\frac{2\beta}{r^{2}}\\
R^{\mu\nu}R_{\mu\nu}
=&\frac{2r_{0}^{2}(1-\beta)^{2}+[(1-\beta)r_{0}+2\beta r]^{2}}{2r^{6}}\\
 R_{\mu\nu\kappa\lambda}R^{\mu\nu\kappa\lambda}
=& \frac{2(1-\beta)^{2}r_{0}^{2}+4[(1-\beta)r_{0}+\beta r]^{2}}{r^{6}}.
\end{split}
\end{equation}
Therefore, in order to guarantee the positivity of the function $f$, we must choose one of the following forms for $f$:
\begin{eqnarray}\label{fKretsch}
f_1&=&\xi R=\frac{2\xi\beta}{r^{2}}, \\
f_2&=&\xi(R^{\mu\nu}R_{\mu\nu})^{1/2}=\frac{\xi\sqrt{2r_{0}^{2}(1-\beta)^{2}+[(1-\beta)r_{0}+2\beta r]^{2}}}{\sqrt{2}r^{3}},\\ f_3&=&\xi(R_{\mu\nu\kappa\lambda}R^{\mu\nu\kappa\lambda})^{1/2}=\frac{\xi\sqrt{2(1-\beta)^{2}r_{0}^{2}+4[(1-\beta)r_{0}+\beta r]^{2}}}{r^{3}}.
\end{eqnarray}
We will first consider the case of Schwarzschild wormholes, that is, $\beta =0$. As we can see $f_1 = 0$ in this case, and so we will just consider $f_2$ and $f_3$ providing for $\beta =0$:
\begin{eqnarray}
f_2 &=& \xi\sqrt{\frac{3}{2}}\frac{r_0}{r^3}, \\
f_3 &=& \xi\sqrt{6}\frac{r_0}{r^3}.
\end{eqnarray}
We can easily see that $f_3 = 2f_2$, and therefore, these cases must have similar behaviors. We first consider $f_2$ to get some conclusions and in the end we plot the graphics for the $f_3$ case. 

In order to verify if the radial energy conditions are satisfied for the Schwarzschild Wormhole, we use $f_2$ in the equations for the state parameters, \eqref{IEQ-tt-s},\eqref{IEQ-rr-s} and \eqref{IEQ-ll-s}, providing for the $\beta=0$ case

\begin{eqnarray}
\kappa\rho &=& \frac{\lambda r_0}{r^6}\left[\frac{15r_0}{2}-6r\right], \\
\kappa p_r &=& \frac{r_0}{r^6}[5\lambda r_0-6\lambda r-r^3], \\
\kappa(\rho+p_r) &=& \frac{r_0}{r^6}\left[\frac{25\lambda r_0}{2}-12\lambda r-r^3\right],
\end{eqnarray}
where just for convenience we define $\lambda = \xi\sqrt{\frac{3}{2}}$. As the multiplicative factor of the above equations are always positive, what will determine the signal of $\rho$, $p_r$ and $\rho + p_r$ are the terms in parentheses. Interestingly, these terms for the energy density $\rho$ is a decreasing linear function of $r$, that is positive for $r<\frac{15}{12}r_0$, while the radial pressure $p_r$ and the sum $\rho+p_r$ are cubic equations having similar behaviors, that is, they are always decreasing functions, starting with positive values and changing sign at their roots.

Let us analyze the behavior of these functions in the wormhole throat. In $r=r_0$ we obtain for the state parameters:

\begin{eqnarray}
\kappa\rho &=& \frac{3\lambda}{2r_0^4}, \\
\kappa p_r &=& -\frac{1}{r_0^4}(\lambda + r_0^2), \\
\kappa(\rho+p_r) &=& \frac{1}{r_0^4}\left(\frac{\lambda}{2}-r_0^2\right).
\end{eqnarray}
We can see that in the throat of a Schwarzschild wormhole the energy density is a positive constant, therefore $\rho > 0$ for all $r_0 > 0$. The sum $\rho+p_r$ is positive for $r_0 < \sqrt{\frac{\lambda}{2}}$ and therefore the Null Energy Condition (NEC) and the Weak Energy Condition (WEC) are satisfied for these values of $r_0$. Furthermore, the radial pressure $p_r$ is always negative and crescent, becoming null in the limit $r_0 \rightarrow \infty$, and it is easy to show that we have $\rho > |p_r|$ for $r_0 < \sqrt{\frac{\lambda}{2}}$. Therefore, the Dominant Energy Condition (DEC) is also satisfied for these values of $r_0$. Thus, the radial energy conditions are satisfied on the throat of a Schwarzschild wormhole, depending on the relation between $r_0$ and $\lambda$:

$$
\begin{cases}\label{energyconditionsr0}
\mbox{Null:}\quad\quad\quad\rho+p_{r}>0 & \mbox{if}\quad r_{0}<\sqrt{\frac{\lambda}{2}},\\
\mbox{Weak:}\quad\rho\geq 0,\rho+p_{r}\geq 0 & \mbox{if}\quad r_{0}<\sqrt{\frac{\lambda}{2}}\\
\mbox{Dominant:}\rho\geq 0,\rho\geq |p_{r}| & \mbox{if}\quad r_{0}<\sqrt{\frac{\lambda}{2}}.
\end{cases}
$$
So, we see that for the Schwarzschild wormhole the Null,Weak and Dominant energy conditions are satisfied nearby the throat in the ASG context for $r_0 < \sqrt{\frac{\lambda}{2}}$. This is in contrast to the results predicted by the usual theory, as we can see taking the limit $\xi \rightarrow 0$, where none of these conditions are satisfied. Now, we analyze what kind of matter source threading the wormhole is allowed in order to ensure that the energy conditions are satisfied nearby the throat, that is, considering $r_0<\sqrt{\frac{\lambda}{2}}$.

Let us start by analyzing what kind of cosmological matter we should have at the throat of a Schwarzschild wormhole if all the energy conditions satisfied. We do this calculating the state parameter $\omega(r) = p_r/\rho$. Evaluating $\omega$ in $r=r_0$ provides:

\begin{equation}
\omega = -\frac{2}{3}\left(1+\frac{r_0^2}{\lambda}\right),
\end{equation}
and we analyze this as a function of $r_0$. Note that the term in parenthenses is always positive, and therefore we always have $\omega < 0$. This feature is quite remarkable because $\omega \rightarrow \infty$ for a Schwarzschild wormhole in the General Relativity context, while in the ASG scenario we obtained a finite $\omega$ in $r_0$. Furthermore we can easily show that for $r_0<\sqrt{\frac{\lambda}{2}}$ we obtain $-1<\omega<-1/3$. We get:

\begin{equation}
\begin{cases}\label{closer_0}
\mbox{Quintessence:}-1<\omega<-1/3 & \mbox{if}\quad r_{0}<\sqrt{\frac{\lambda}{2}},\\
\mbox{Phantom:}\quad\quad\quad\quad\omega<-1 & \mbox{if}\quad r_0>\sqrt{\frac{\lambda}{2}}.\\
\end{cases}
\end{equation}
Therefore, in order to have all the energy conditions satisfied, the throat must be sourced by a Quintessencial fluid.

Now, considering $r_0 <\sqrt{\frac{\lambda}{2}}$, that is, the case where the throat is sourced by quintessencial fluid, we analyze what kind of matter we must have region by region, analyzing the signal of the $\omega(r)=p_r/\rho$ :

\begin{equation}
\omega(r) = \frac{5\lambda r_0-6\lambda r-r^3}{\frac{15\lambda r_0}{2}-6\lambda r},
\end{equation}
where $\lambda = \xi\sqrt{\frac{3}{2}}$ and $r_0 < \sqrt{\frac{\lambda}{2}}$. This function has an asymptote in $r = \frac{15r_0}{12}$ and we can see that for the region $r>\frac{15r_0}{12}$ we always have exotic matter with $\omega >1$. For $r_{-}<r<\frac{15r_0}{12}$, we have $\omega<-1$, a region with Phantom-like exotic matter, where $r=r_-$ is the value of $r$ whose $\omega = -1$, or equivalently, when $\rho + p_r =0$, and in terms of $\lambda$ and $r_0<\sqrt{\frac{\lambda}{2}}$ is given by:

\begin{equation}
r_- = \frac{6,3496 \lambda }{\sqrt[3]{\sqrt{625r_0^2 \lambda ^2+1024\lambda ^3}-25r_0\lambda }}-0,629961 \sqrt[3]{\sqrt{625r_0^2 \lambda ^2+1024\lambda ^3}-25r_0 \lambda }.
\end{equation}
Finally, for $r_0<r<r_-$ we can have a region with quintessencial fluid, that is, a exotic matter with $-1<\omega<-1/3$. Therefore, we have for the solutions region by region for the Schwarzschild Wormhole:

\begin{equation}
\begin{cases}\label{regionsr}
\mbox{Quintessence:}-1<\omega<-1/3 & \mbox{if}\quad r_{0}<r<r_-,\\
\mbox{Phantom:}\quad\quad\quad\quad\omega<-1 & \mbox{if}\quad r_-<r<\frac{15r_0}{12},\\
\mbox{Other Exotic Matter:}\quad\quad\quad\quad\quad\omega>1 & \mbox{if}\quad r>\frac{15r_0}{12}.
\end{cases}
\end{equation}
As we can see, the quantum improvement leads to the inevitable presence of phantom-like matter regions as source to generate the wormhole spacetime. These results are very similar with the Ellis-Bronnikov wormhole \cite{Ellis:1973yv,Bronnikov:1973fh} case, which was studied by Alencar et al. \cite{Alencar:2021enh}. They found that for the ASG context the radial energy conditions are satisfied in the throat for a range of values of $r_0$, and for this is necessary a exotic matter, like quintessencial fluid threading the wormhole throat. Furthermore, we also necessarily have the presence of regions with phantom-like matter. This is exactly what happens with Schwarzschild wormholes in the ASG scenario.

\begin{figure}[htb]
\centering
\includegraphics[width=0.6\textwidth]{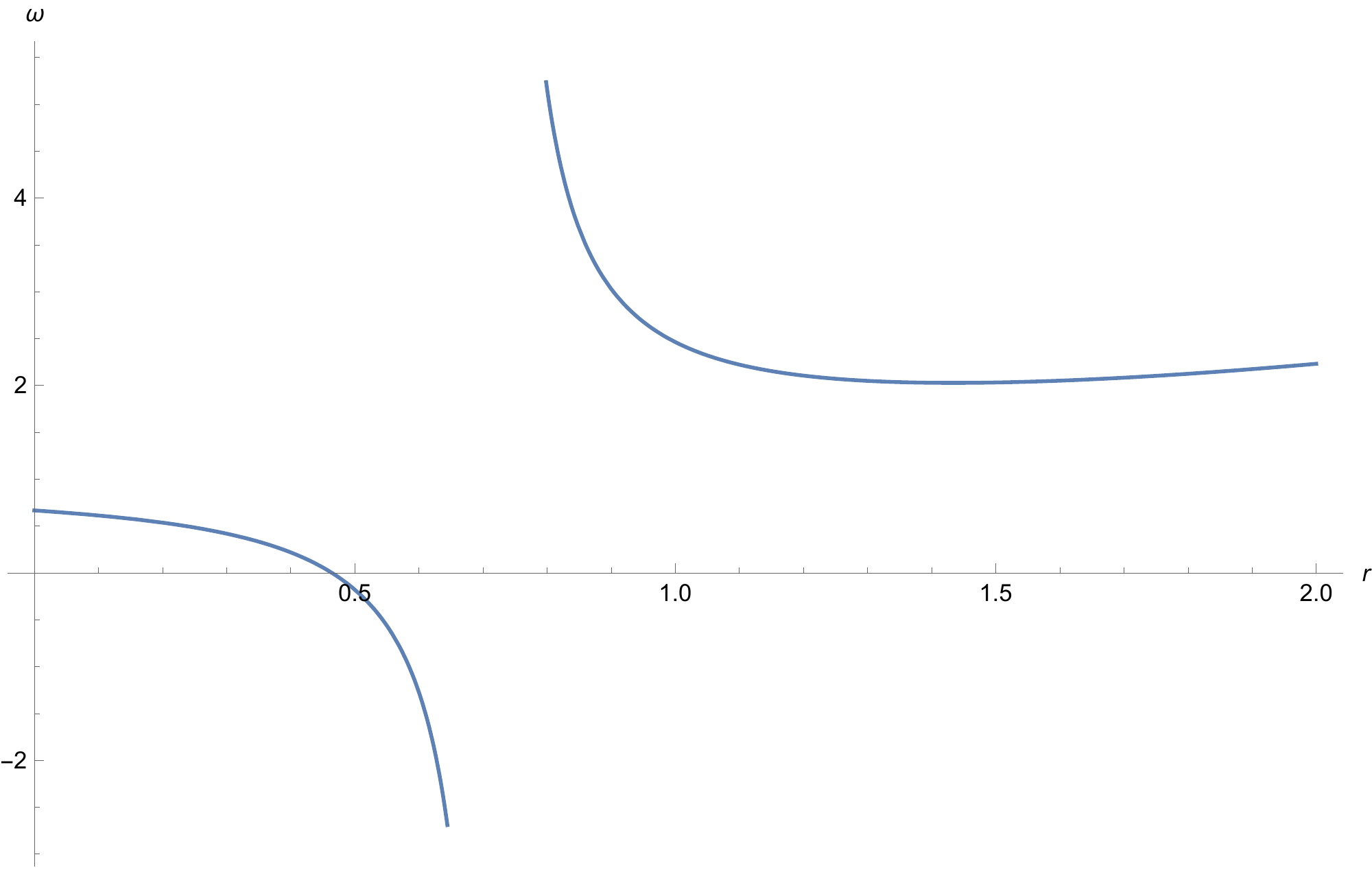}
\caption{Plot of $\omega(r)$ for a zero-tidal Schwarzschild wormhole($\beta=0$) in the ASG context for $\lambda=1$($\xi = \sqrt{2/3}$) and $r_0^2=1/3$.}
\label{omega-SW-ASG}
\end{figure}

The behavior of $\omega$ for $\lambda=1$ and $r_0^2 = 1/3$ can be seen in the figure \ref{omega-SW-ASG}. In this case we have a small region containing quintessencial fluid for $r > r_0=1/\sqrt{3}$ and becomes a Phantom region in $r = r_- \approx 0,587$ until the moment when $\omega$ diverges in $r \approx 0,721$, and for $r > 0,721$ we have exotic matter with $\omega > 1$.

\begin{figure}[htb]
\centering
\includegraphics[width=0.6\textwidth]{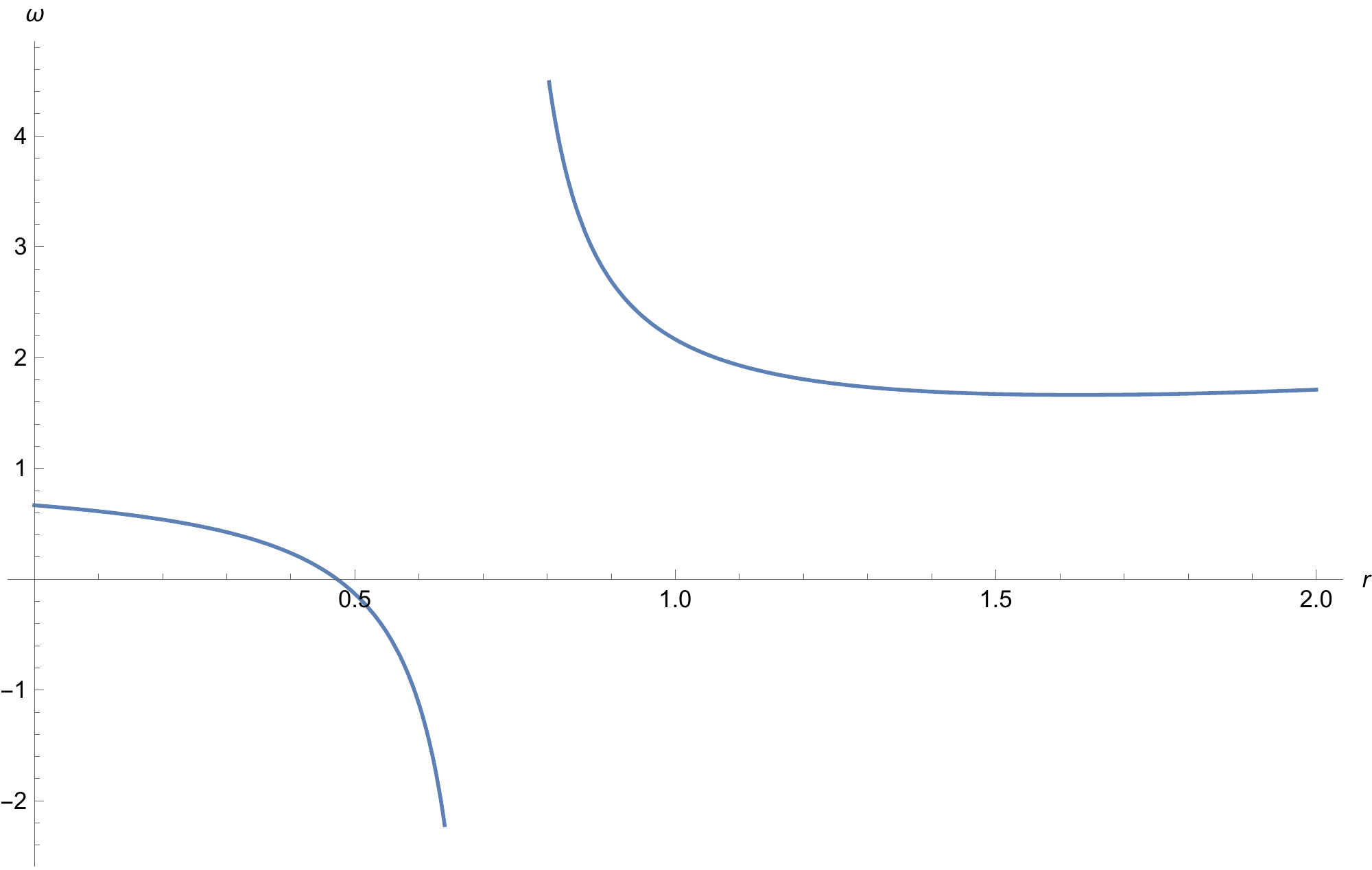}
\caption{Plot of $\omega(r)$ for a zero-tidal Schwarzschild wormhole($\beta=0$) using the Kretschmann scalar for anti-screening function with $\lambda=1$($\xi = \sqrt{2/3}$) and $r_0^2=1/3$.}
\label{omega-SW-Kretschmann}
\end{figure}

As the case of the Kretschmann scalar has basically the same behavior as the squared ricci tensor case, we only show the plot of the expression for $\omega$, which can be seen in the figure \ref{omega-SW-Kretschmann}, where we have essentially the same behavior as in the previous case.

\section{Schwarzschild-like Wormhole solution in ASG}
Now we turn our attention to the Schwarzschild-like wormhole case, that is, considering $\beta \neq 0$. For this, we consider the Ricci scalar model to the anti-screening function $f = f_1 = \xi R$. In order to ensure the condition $f_1 > 0$ we have to separate $f_1$ in two forms depending on the signal of $\beta$:

\begin{equation}
\label{function-SWL}
f_1 = 
\left\{
\begin{array}{cc}
-2\xi\beta/r^{2}\quad\quad\textrm{if}\quad\beta<0  \\
 2\xi\beta/r^{2}\quad\quad\textrm{if}\quad 0<\beta<1,
\end{array}
\right.
\end{equation}
that is, we consider the minus sign when $\beta < 0$ and the plus sign when $0<\beta<1$ to guarantee the positivity of the anti-screening function. Therefore, once we get the modified state equations for one of the cases, just do the prescription $\xi \rightarrow -\xi$ to obtain these quantities for the other case.

Now we will analyze the radial conditions of energy of the Schwarzschild-like Wormhole. Those conditions are verified from the substitution of Eq. (\ref{function-SWL}) into Eqs. (\ref{IEQ-tt-s}) and (\ref{IEQ-rr-s}). We get for the $0<\beta<1$ case:

\begin{eqnarray}
\kappa\rho&=&\frac{\beta}{r^{5}}\left[r^{3}-2\xi(2-3\beta)r+6\xi r_{0}(1-\beta)\right],\label{rhoellis1}\\
\kappa p_r&=&\frac{\beta}{r^{5}}\left[-r^{3}+r_{0}\left(1-1/\beta\right)r^{2}-2\xi(4-3\beta)r+6\xi r_{0}(1-\beta)\right],\label{prellis1}\\
\kappa(\rho+p_{r})&=&\frac{\beta}{r^{5}}\left[r_{0}(1-1/\beta)r^{2}-12\xi(1-\beta)r+12\xi r_{0}(1-\beta)\right],\label{plellis1}
\end{eqnarray}
and the case $\beta<0$ is reached making the prescription $\xi \rightarrow -\xi$ providing:

\begin{eqnarray}
\kappa\rho&=&\frac{\beta}{r^{5}}\left[r^{3}+2\xi(2-3\beta)r-6\xi r_{0}(1-\beta)\right],\label{rhoellis2}\\
\kappa p_r&=&\frac{\beta}{r^{5}}\left[-r^{3}+r_{0}\left(1-1/\beta\right)r^{2}+2\xi(4-3\beta)r-6\xi r_{0}(1-\beta)\right],\label{prellis2}\\
\kappa(\rho+p_{r})&=&\frac{\beta}{r^{5}}\left[r_{0}(1-1/\beta)r^{2}+12\xi(1-\beta)r-12\xi r_{0}(1-\beta)\right].\label{plellis2}
\end{eqnarray}
What will determine the signal of the $\rho$,$p_{r}$ and $\rho + p_{r}$ are the terms between parentheses. The terms in $\rho$ and $p_{r}$ are cubic equations that have just one real root and the term for $\rho + p_{r}$ is a quadratic equation. We can readily see that these expressions are all positive for $r=0$ if $0<\beta<1$ and are all negative if $\beta<0$. Therefore, close to $r=0$ we can say that they are all positive, satisfying the energy conditions for both conditions of $\beta$.

However, as in the $\beta = 0$ case, we will first analyze the signal of the quantities $\rho$, $p_r$, and $\rho+p_r$, in order to verify if the null ($\rho+p_r\geq 0$), weak ($\rho\geq 0$, $\rho+p_r\geq 0$) and dominant ($\rho\geq 0$, $\rho\geq|p_r|$) energy conditions are satisfied nearby the throat. For this, we consider the equations \eqref{rhoellis1},\eqref{prellis1},\eqref{plellis1} in $r=r_{0}$ providing for the $0<\beta<1$ case:

\begin{eqnarray}
\kappa\rho&=&\frac{\beta}{r_{0}^{4}}\left(2\xi+r_{0}^{2}\right),\label{rhor0}\\
\kappa p_r&=&-\frac{\beta}{r_{0}^{4}}\left(\frac{r_{0}^{2}}{\beta}+2\xi\right),\label{prr0}\\
\kappa(\rho+p_{r})&=&\frac{(\beta - 1)}{r_{0}^{2}}.\label{plr0}
\end{eqnarray}

Therefore we arrive at some general conclusions in this case. We see that $p_r$ must be a crescent function of $r_0$ and it is always negative, such that $p_r \rightarrow 0$ in the limit $r_0 \rightarrow \infty$. On other hand, we have $\rho>0$ for all $r_{0}>0$ since $\rho$ is formed only for positive terms. Finally, note that $\rho+p_r<0$ for all $r_{0}>0$ because of the condition $\beta<1$. Due to the $\beta^{-1}$ term in $p_{r}$ we can note that $\rho \leq |p_{r}|$. Thus, we see that none of the Energy Conditions are not satisfied for $r_{0}>0$ in the $0<\beta<1$ case.

Doing $\xi \rightarrow -\xi$ we obtain the equations for the state parameters nearby the wormhole throat for the case $\beta <0$ providing:

\begin{eqnarray}
\kappa\rho&=&\frac{\beta}{r_{0}^{4}}\left(-2\xi+r_{0}^{2}\right),\label{rhor0}\\
\kappa p_r&=&\frac{\beta}{r_{0}^{4}}\left(-\frac{r_{0}^{2}}{\beta}+2\xi\right),\label{prr0}\\
\kappa(\rho+p_{r})&=&\frac{(\beta - 1)}{r_{0}^{2}}.\label{plr0}
\end{eqnarray}

As we can see $\rho + p_{r}$ remains negative for all $r_{0}>0$ and therefore the Null and Weak conditions are not satisfied. For $\rho > 0$ we must have $r_{0}<\sqrt{2\xi}$ and we can note that $p_{r}<0$ for all $r_{0}>0$. However, $\rho \leq |p_{r}|$ in $r_{0}<\sqrt{2\xi}$ and therefore the radial energy conditions are still not satisfied in $\beta<0$ case. 

The results of General Relativity can be readily recovered by doing the limit $\xi \rightarrow 0$, as we can see in the Ref\cite{Cataldo:2017ard}. In the usual case the radial energy conditions are not satisfied in the wormhole throat and this feature remains in the context of ASG theory, at least for the Ricci scalar case.

Now, we will investigate the presence of cosmologic matter in the Schwarzschild-like wormhole in ASG theory by evaluating the state parameter $\omega(r)=p_r/\rho$.  Let us first see what kind of matter is allowed at our wormhole throat. For this we have that for the $0<\beta<1$ case:

$$
\omega=\frac{-r_0^2/\beta-2\xi}{r_0^2+2\xi},
$$
and we can analyze this as a function of $r_0$. We can see immediately that the $\omega$ is always negative in this case, and therefore the throat can not be formed by ordinary matter$(0<\omega<1)$ or other exotic Matter with $\omega >1$. Furthermore it is easy to note that $\omega<-1$ for all $r_{0}>0$ and we conclude that nearby the throat we must necessarily have a Phantom fluid, if $0<\beta<1$. Again, this is the same result found in the usual case, as we can see performing the limit $\xi \rightarrow 0$, providing $\omega = -1/\beta$. 

Doing $\xi \rightarrow -\xi$ we obtain the state parameter in $r=r_0$ for the $\beta<0$ case:

$$
\omega = \frac{-r_0^2/\beta+2\xi}{r_0^2-2\xi}.
$$
It is easy to show that for $r_{0}<\sqrt{2\xi}$ we have always $\omega<-1$, for all $\beta < 0$, characterizing a wormhole with phantom-like matter on your throat. But, this time, we have the possibility of finding ordinary matter($0<\omega<1$) in the throat, as long as the values of $\beta$ are also restricted to $\beta<-1$. We get easily:

\newcommand{\bb}[1]{\raisebox{-2ex}[0pt][0pt]{\shortstack{#1}}}
\renewcommand{\arraystretch}{1.9}
\begin{tabular}{c|c|c}\hline\label{closer_0}
Phantom:&$\omega<-1$& if  $r_0<\sqrt{2\xi}$ for all $\beta <0$\\\hline
\bb{Other Exotic Matter:}&\bb{$\omega>1$}&if $\sqrt{2\xi}<r_0$  with $-1<\beta<0$ \\\cline{3-3}
&&$\sqrt{2\xi}<r_0< \sqrt{\frac{4\xi}{1+1/\beta}}$ with $\beta < -1$ \\\hline
Ordinary Matter:&$0<\omega<1$& if $\quad r_0 > \sqrt{\frac{4\xi}{1+1/\beta}}$ with  $\beta <-1$ \\\hline
\end{tabular}
\renewcommand{\arraystretch}{1}

Therefore, in the $\beta < 0$ case we have the possibility of the throat being formed by exotic matter of the Phantom-type($\omega<-1$), ordinary Matter($0<\omega<1)$ or other exotic Matter with $\omega >1$ depending of the values of $r_0$ and $\beta$. This result differs from the usual case because there is the possibility of the throat be formed by a exotic matter Phantom-like, while in General Relativity we always have $\omega = -1/\beta > 0$ in $r=r_0$, that is, ordinary matter($0<\omega<1$) if $\beta <-1$ or exotic matter with $\omega > 1$ if $-1<\beta<0$ nearby the throat.

\begin{figure}[htb]
\centering
\includegraphics[width=0.6\textwidth]{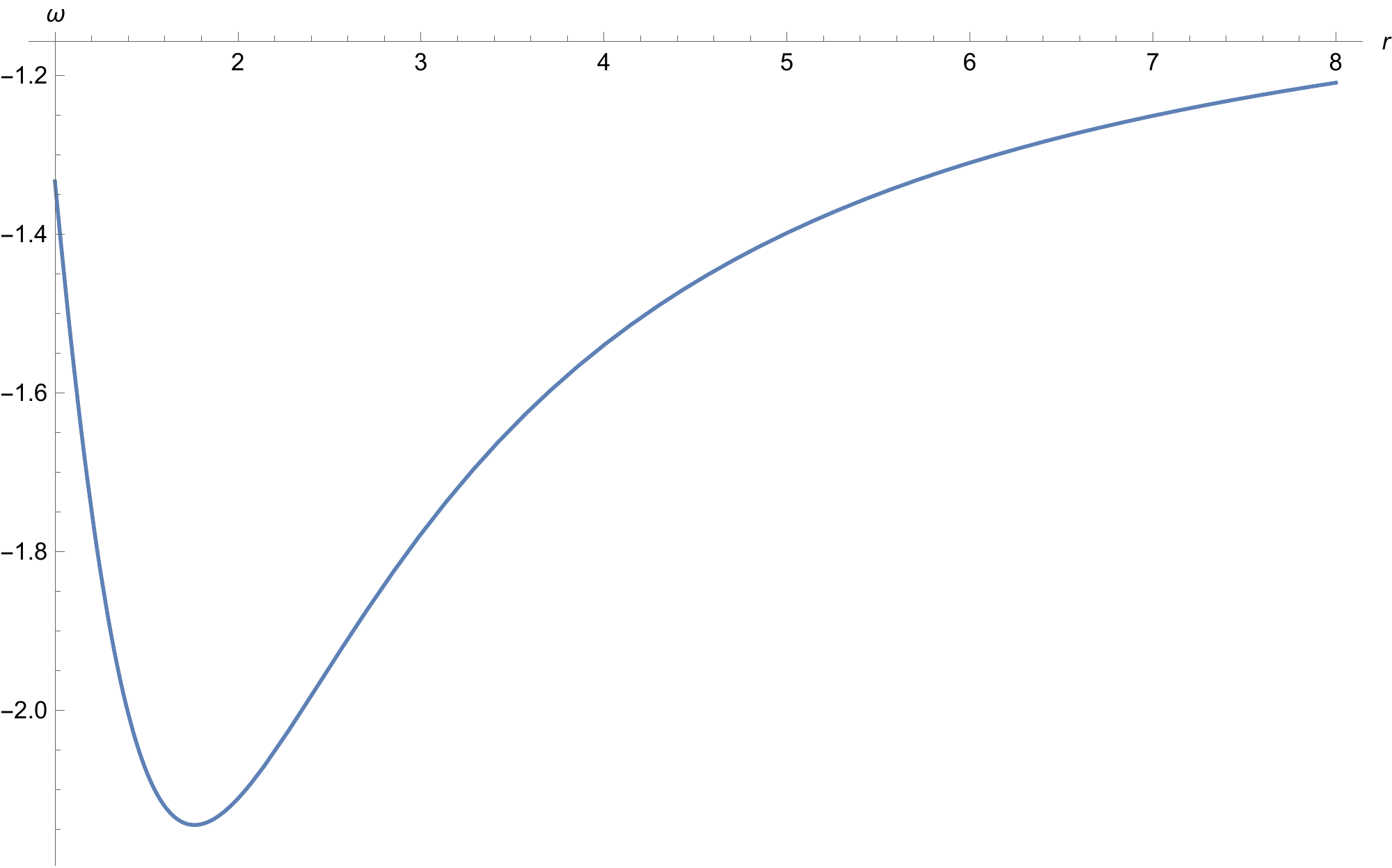}
\caption{Plot of $\omega(r)$ for a zero-tidal Schwarzschild-like wormhole for $r_0 = 1 < r$, $\xi = 1$ and $\beta = 1/2$, in Planckian units.}
\label{omega-SWL-bpositive}
\end{figure}

Finally, we analyze what type of matter we must have for all the Schwarzschild-like wormhole spacetime. For this we have for the $0<\beta<1$ case:

$$
\omega (r) = \frac{p_r(r)}{\rho(r)} = \frac{-r^3+r_0(1-1/\beta)r^2-2\xi(4-3\beta)r+6\xi r_{0}(1-\beta)}{r^{3}-2\xi(2-3\beta)r+6\xi r_{0}(1-\beta)}.
$$
As noted earlier, in this case we always have Phantom fluid at the wormhole throat for all $r_0>0$. Note that the denominator is always a crescent and positive function for $r>r_{0}>0$, and therefore, what will determine the signal of the expression above is the numerator. It can be shown that the numerator is a decrescent function, starting positive and changing its sign at its root. However, whatever the value of $r_0$, its root will always be smaller than $r_0$. Therefore, we can not have regions with ordinary Matter($0<\omega<1$) or other exotic matter with $\omega >1$ . Furthermore, as long as $r_0 < r$, we always have $\rho \leq |p_{r}|$, and so $\omega < -1$ for all $0<r_0<r$. Therefore, for Schwarzschild-like wormholes with a lack of solid angle($0<\beta<1$) we have necessarily a Phantom wormhole. We show this features in Fig. \ref{omega-SWL-bpositive}, where we can see  that the entire region of the wormhole must be filled with phantom-like matter, with $\omega \rightarrow -1$ when $r \rightarrow \infty$.

Doing the exchange $\xi\rightarrow -\xi$, we have the state parameter $\omega (r)$ for the $\beta <0$ case:

$$
\omega (r) = \frac{-r^3+r_0(1-1/\beta)r^2+2\xi(4-3\beta)r-6\xi r_{0}(1-\beta)}{r^{3}+2\xi(2-3\beta)r-6\xi r_{0}(1-\beta)}.
$$
Let us consider the case where the throat is sourced by phantom fluid, that is, $r_0 < \sqrt{2\xi}$ for all $\beta <0$. In this case the denominator is a crescent function, starting negative for $0<r_0<r$ and changing its signal in $r=a$, where $a$ is the root of the denominator, which in terms of $\xi$ and $r_0<\sqrt{2\xi}$ is given by:

\begin{equation}
a = \frac{2\sqrt[3]{3}(-2+3\beta)\xi+\left[27(1-\beta)\xi r_0+\sqrt{24(2-3\beta)^{3}\xi^{3}+729(1-\beta)^{2}\xi r_0^{2}}\right]^{2/3}}{3^{2/3}\left[27(1-\beta)\xi r_0+\sqrt{24(2-3\beta)^{3}\xi^{3}+729(1-\beta)^{2}\xi r_0^{2}}\right]^{1/3}}.
\end{equation}
Therefore, the state parameter has an asymptote in this case, separating $\omega$ in two regions. Furthermore, the numerator has a positive region for $r_0<r<b$, where $r = b$ is the value of its root and becomes negative for $r>b$. 

\begin{figure}[!htb]
\centering
\includegraphics[width=0.6\textwidth]{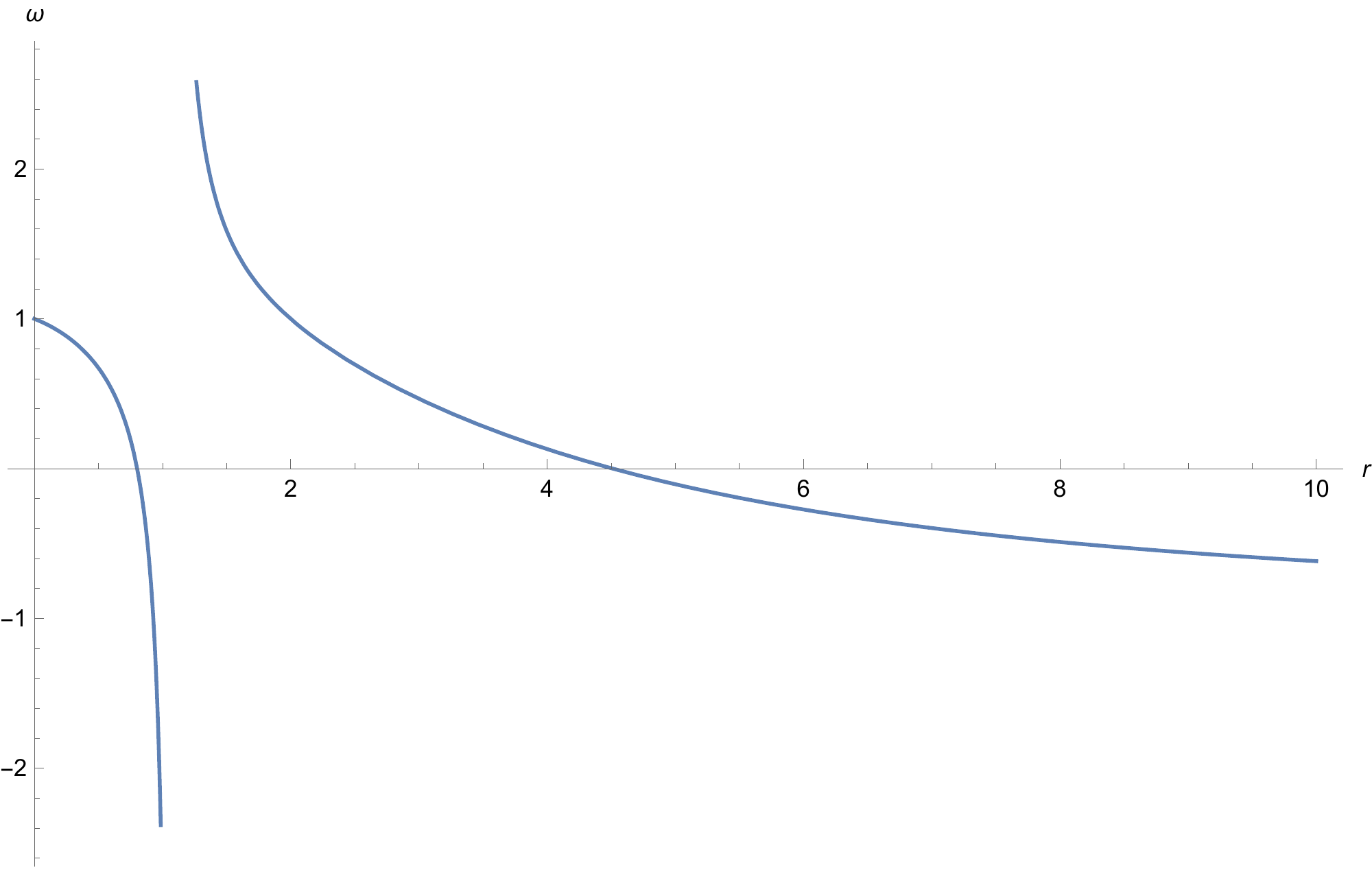}
\caption{Plot of $\omega(r)$ for a zero-tidal Schwarzschild-like wormhole with an excess of solid angle ($\beta<0$) for $\xi=1$, $r_0=1$, and $\beta = -1$.}
\label{omega-SW-negativo}
\end{figure}

It is easy to show that we have $\omega<-1$ for $r_0<r<a$, more Phantom matter after the throat region. When $r>a$ there is a smooth decay of $\omega$, with $\omega>1$ for $a<r<r_+$, where $r=r_+$ is the value of the radial coordinate when $\omega = 1$, given by:

\begin{equation}
r_+ = \frac{r_0(1-1/\beta)+\sqrt{r_0^2(1-1/\beta)^2+32\xi}}{4}.
\end{equation}
Therefore we must have $0<\omega<1$ for $r_+<r<b$, where $\omega=0$ in $r=b$. Interestingly, if the throat of a Schwarzschild-like wormhole with excess of solid angle ($\beta <0$) is formed by Phantom matter we have the possibility of a region with nonexotic matter($0<\omega<1$). Next, $\omega$ becomes negative again for $r>b$ and keeps decaying until reach $-1/3$ for some $r=c$, and tending to $-1$ for $r_0 \rightarrow \infty$. Therefore,  we must have a large presence of quintessencial fluid for $r>c$, and we easily get:

\begin{equation}
\begin{cases}\label{regionsr}
\mbox{Phantom:}\quad\quad\quad\quad \omega<-1 & \mbox{if}\quad , r_{0}<r<a, \\
\mbox{Other exotic matter:}\quad\quad\quad\quad\omega>1 & \mbox{if}\quad a<r<r_+,\\
\mbox{Ordinary Matter:}\quad\quad\quad\quad\quad0<\omega<1 & \mbox{if}\quad r_+<r<b, \\
\mbox{Quintessence:}\quad\quad\quad\quad -1/3<\omega<-1 & \mbox{if}\quad r>c.
\end{cases}
\end{equation}
The features above for $\omega$ in this case, for $\xi = 1$, $r_0=1$ and $\beta = -1$, is plotted in Fig. \ref{omega-SW-negativo}. We have phantom-like matter for $r_0=1<r<1,08$ followed by a region with exotic matter with $\omega > 1$ until reaches the region where there is non-exotic matter($0<\omega<1$) for $2<r<4,51$. Finally, the final region is sourced by quintessence-like matter for $r>6,45$.

\newpage

\section{Conclusions}
In this paper, we have considered zero-tidal Schwarzschild-like wormholes in the Asymptotically Safety Gravity (ASG) scenario to improve quantum mechanically the General Relativity. For this, it was considered  models that use Ricci scalar, squared Ricci tensor and Kretschmann scalar to make functional renormalization group improvement to describe ASG.

In this context, we have first analyzed the particular case of a Schwarzschild wormhole, which is asymptotically flat, using the squared Ricci tensor and the Kretschmann scalar. Both models provide identical improvement in this case, since they are proportional. In this case we have shown that the radial energy conditions are satisfied nearby its throat, as long as we have $r_0 < \sqrt{\frac{\lambda}{2}}$, where we defined $\lambda = \sqrt{\frac{3}{2}}\xi$. Furthermore, analyzing the signal of the state parameter $\omega = p_r/\rho$ as a function of $r_0$, we have determined that the quantum corrections imply that regions nearby the throat must be sourced by quintessence-like matter($-1<\omega<-1/3$) if $r_0 < \sqrt{\frac{\lambda}{2}}$ or phantom-like matter ($\omega<-1$) if $r_0 > \sqrt{\frac{\lambda}{2}}$, where we see that if the radial energy conditions are satisfied the throat must necessarily be formed by quintessence-like matter. These results are quite different from those obtained by General Relativity, since $\omega(r_0)$ must diverge in the usual case.

Next, we studied the types of sources allowed along the spacetime generated by the Schwarzschild wormhole in this quantum modified gravity, if the energy conditions are satisfied in the wormhole throat. With these considerations, we show that the wormhole can be separated in three regions, as we move away from the throat region. First, we found a small extension containing quintessence-like matter. Proceeding, we enter in a region that is sourced by phantom-like matter, finishing with a region sourced by exotic matter with $\omega > 1$. These features have been shown in Fig. \ref{omega-SW-ASG} for the squared Ricci tensor model, and in Fig. \ref{omega-SW-Kretschmann} for the Kretschmann model, where we basically have the same behavior. These results are analogous to those found for other asymptotically flat wormholes, such as Ellis-Bronnikov one, studied in \cite{Alencar:2021enh}, where the authors showed that is the possible that the radial energy conditions are satisfied nearby the throat, but the presence of exotic matter, such as phantom or quintessence, is always necessary.

Thereafter, we repeat the same procedure described above for the more general case of Schwarzschild-like wormholes, whose new asymptotic form is deformed by an increase or decrease in the solid angle of the sphere, depending on the sign of a constant $\beta$. We have seen that, just as in the context of General Relativity, the radial energy conditions are not satisfied nearby the throat in both cases. When analyzing the state parameter $\omega$ as a function of $r_0$, we shown that, when there is a lack of solid angle, we necessarily have phantom-like matter for regions close to the throat, and when there is an excess of solid angle, the throat can be formed by exotic matter with $\omega > 1$, ordinary matter($0<\omega<1$) or phantom-like matter, depending on the values of $r_0$ and $\beta$. These results are also obtained in the usual case, except for the latter, where the possibility of the throat being formed by phantom-like matter arises due to the quantum effects considered here. 

Finally, we study the types of matter allowed along the spacetime determined by the Schwarzschild-like wormhole. We shown that for a lack of solid angle, beyond its throat, we necessarily have phantom-like matter for all the wormhole extension, characterizing a Phantom wormhole, as we saw in Fig. \ref{omega-SWL-bpositive}. For an excess of a solid angle, we considered that the throat is formed by phantom-like matter, induced by quantum corrections, and we see that in this case the wormhole can be divided into four regions. First we have a small extension containing phantom-like matter followed by a region with exotic matter with $\omega >1$. Then, we come into a region sourced by ordinary matter, 
and the final region is formed by quintessence-like matter.

Therefore, we can conclude that the quantum improvement provided the possibility of matter satisfying the radial energy conditions nearby the throat only in the asymptotically flat case, although in this case it is necessary quintessence-like matter. This implies the presence of very exotic matter throughout the modified space-time, such as phantom-like matter. In case that have a deformation in the asymptotic form of the metric the radial energy conditions are not satisfied nearby the throat, but, the quantum corrections imply one more time the possibility of having regions with phantom-like matter, as was found for other wormholes in the ASG context, such as Ellis-Bronnikov wormhole. These results supports the idea that the presence of a phantom fluid are unavoidable in the context of Asymptotically Safe Gravity, and perhaps in other quantum gravity models.
\vspace{6pt} 




\funding{The authors would like to thank Conselho Nacional de Desenvolvimento Cient\'{i}fico e Tecnol\'{o}gico (CNPq), Funda\c{c}\~{a}o Cearense de Apoio ao Desenvolvimento Cient\'{\i}fico e Tecnol\'{o}gico (FUNCAP) and Coordena\c c\~{a}o de Aperfei\c coamento de Pessoal de N\'{i}vel Superior - Brasil (CAPES) for finantial support.}

\conflictsofinterest{The authors declare no conflict of interest.} 




\end{paracol}
\reftitle{References}

\end{document}